# Breaking a fully Balanced ASIC Coprocessor Implementing Complete Addition Formulas on Weierstrass Elliptic Curves

Ievgen Kabin[1], Zoya Dyka[1], Dan Klann[1], Nele Mentens[2,3], Lejla Batina[4] and Peter Langendoerfer[1,5]

[1]IHP – Leibniz-Institut für innovative Mikroelektronik, Frankfurt (Oder), Germany
[2]ES&S and imec-COSIC/ESAT, KU Leuven, Belgium
[3]LIACS, Leiden University, Leiden, The Netherlands
[4]Digital Security Group, Radboud University, Nijmegen, The Netherlands
[5]BTU Cottbus-Senftenberg, Cottbus, Germany
{kabin, dyka, klann, langendoerfer}@ihp-microelectronics.com, nele.mentens@kuleuven.be, lejla@cs.ru.nl

*Abstract*— In this paper we report on the results of selected horizontal SCA attacks against two open-source designs that implement hardware accelerators for elliptic curve cryptography. Both designs use the complete addition formula to make the point addition and point doubling operations indistinguishable. One of the designs uses in addition means to randomize the operation sequence as a countermeasure. We used *the comparison to the mean* and an automated SPA to attack both designs. Despite all these countermeasures, we were able to extract the keys processed with a correctness of 100%.

## I. INTRODUCTION

Cryptographic protocols using Elliptic Curves (EC) over finite fields are world-wide known approaches for the generation and verification [1] of digital signatures as well as for mutual authentication. EC cryptographic operations are time and energy expensive, but significantly faster than RSA [2]. Additionally, Elliptic Curve Cryptography (ECC) uses cryptographic keys that are significantly shorter than by RSA while providing the same level of security. This reduces the time and energy required for sending and receiving the messages. These features make ECC very attractive for resource-constrained devices that require not only a high level of security but also low-power real-time communication and data processing. The application areas for which this is of importance are the Internet of Things (IoT), autonomous driving, e-health, Industry 4.0 and many other applications.

The main operation of ECC-based protocols is the EC point multiplication denoted as $kP$. $P$ denotes an EC point and $k$ is a scalar, i.e. a long binary number. The security requirements define the length of the scalar $k$. Nowadays the recommended lengths of the scalar are 256 bits for an EC over a prime field and 283 bits for an EC over an extended binary field. In EC-based authentication protocols a device uses its private key as a scalar $k$ performing $kP$ to confirm its identity. The goal of an attacker is to reveal the private key. For generating a digital signature using ECDSA [1] a randomly generated long number is used as the scalar $k$. If an attacker can reveal this random number, he can easily calculate the private key used for generating the signature. During the execution of cryptographic operations with the private key, the energy consumed depends on the processed inputs and on the private key. If the attacker can measure the current drawn from the power supply or the electromagnetic emanation of an unprotected device, the used private key can be revealed by statistical analysis of the measured trace(s). Such attacks – i.e. Side Channel Analysis (SCA) attacks – are often classified into vertical and horizontal attacks. For vertical attacks at least two traces have to be measured and analysed. Usually attackers need to analyse differences in hundreds of traces to reveal a key bit. Attacks performing analysis of a single trace are horizontal attacks. The analysis is classified as simple analysis if the key can be revealed via visual inspection of the measured trace. If statistical methods are used the analysis is classified as differential (or correlation) analysis. The main idea of the analysis is to distinguish parts corresponding to the processing of key bits '1' from parts corresponding to the key bits '0' in the measured trace. The $kP$ operation is defined as adding the point $P$ $k$-times to itself and can be calculated as a sequence of EC point additions (denoted as $P+Q$) and point doublings (denoted as $2P$). When processing a key bit '0' only a point doubling has to be calculated. When processing a key bit '1' two EC point operations – a point doubling and a point addition – have to be performed. The formula for point addition differs from the formula for point doubling which makes the power profiles of these operations distinguishable from each other. Depending on the implemented algorithm the differences in the power profiles can easily be seen (simple SCA) or a complex statistical analysis is required to extract the key. The goal of designers is to make the processing of key bits '1' indistinguishable from the processing of key bits '0', i.e. the trace of the implemented algorithm has to be independent of the processed key.

The Montgomery ladder is a well-known algorithm for calculating $kP$ [3]. The algorithm is a bitwise processing of the secret scalar $k$ (further denoted here as the key) from its most significant bit (MSB) to its least significant bit (LSB), i.e. from left to right, see Algorithm 1.

| Algorithm 1: Montgomery ladder [3] |
|---|
| **Input**: $P$, $k=k_{l-1} k_{l-2} ... k_0$, with $k_{l-1}=1$ |
| 1. $Q_0 \leftarrow P$; $Q_1 \leftarrow 2P$; |
| 2. **for** $i = (l-2)$ **downto** $0$ |
| 3.    **if** $k_i = 1$ **then**     $Q_0 \leftarrow Q_0+Q_1$ ; $Q_1 \leftarrow 2Q_1$ |
| 4.    **else**                  $Q_1 \leftarrow Q_0+Q_1$ ; $Q_0 \leftarrow 2Q_0$ |
| 5. **end for** |
| **Output**: $Q_0 = k \cdot P$ |

In the literature the Montgomery ladder is considered to be resistant against simple SCA attacks that are a kind of horizontal attacks. This assessment is based on the fact that it is a balanced algorithm, i.e. the number and the sequence of mathematical operations as well as register operations (reading or storing of the data) do not depend on the value of the processed key bit. Please note that although the sequence of operations, including register operations, is independent of the processed key bit value, the addressing of the registers when processing a key bit value '0' differs from the addressing of the registers when processing a key bit value '1'. Thus, the fact that the addressing of the registers and the storage of data into the registers depend on the key can be exploited to extract the processed key analysing the power consumption or electromagnetic emanation measured during the execution of the Montgomery ladder. The first successful vertical attack exploiting the addressing of the registers was published in 2002 in [4] and is known as address bit differential power analysis (DPA) attack. A horizontal address bit SCA is reported in [5]. Vertical attacks exploiting the energy consumption of storing data into registers are also known as data bit SCA. Thus, the Montgomery $kP$ algorithm is vulnerable to vertical data bit DPA as well as to vertical and horizontal address bit DPA. In [6], the authors proposed a countermeasure against vertical address bit DPA. In [7] randomizing the main loop of the Montgomery $kP$ algorithm was proposed as a countermeasure against vertical data bit SCA. Due to this randomization the addressing of registers no longer depends on the processed key bit value, so it can be a suitable countermeasure against vertical and horizontal address bit SCA also.

A fully balanced ASIC Coprocessor implementing complete addition formulas on Weierstrass Elliptic Curves was reported in [8]. In this paper the two designs proposed in [8] are investigated. The one implements the Montgomery ladder algorithm without randomization of the main loop while the other one implements the randomization. Whereby the EC point operations – the point doubling as well the point addition – were implemented corresponding to a universalized point addition formula with the goal to make their power profiles indistinguishable. The design is open source and implementation details are given in [9]. The resistance of the design against attacks was not evaluated in [8] and to the best of our knowledge also not in any other publication.

In this paper we describe the horizontal attacks we performed against the fully balanced randomized Montgomery $kP$ design implementing complete addition formulas. Our results are that both designs – the one with and the one without randomized main loop are vulnerable to horizontal attacks despite the universalized formula for point addition and point doubling. This paper is structured as follows. In section II we give a short overview of the analyzed designs. Parameters of the designs synthesized for the 250 nm cell library are given in section III. Sections IV describes a horizontal DPA attack using *the comparison to the mean* [10] for revealing the key. Section V shortly describes an automated simple power analysis (SPA) attack. We performed attacks using simulated power traces and evaluated the results using the knowledge of the key. Section VI concludes this work.

## II. BALANCED ELLIPTIC CURVE COPROCESSOR

The investigated designs are area-optimized application-specific integrated circuit (ASIC) implementations of the Montgomery ladder for an elliptic curve point multiplication. We used the open-source VHDL code published by the authors of [8]. The code is available through GitHub [11].

The first open source design is an implementation of the Montgomery ladder based on [3], see for example Algorithm 1. The second open source design implements a random order execution according to the algorithm proposed in [7], here given as Algorithm 2. Please note that the scalar $k$ is denoted as $m$ in Algorithm 2 and in [7], and the length of the scalar $m$ is denoted as $t$. We use this notation in the rest of the paper. In both designs, point doubling and point addition were implemented according to the complete addition law proposed in [12] (see Algorithm 7 in [12]). This leads to identical execution times as well as similar shapes of the power traces for point addition and point doubling since the same addition formulas are used for both operations. Thanks to the fact that the power shapes of the EC point operations are similar, their randomized execution hides the knowledge about the time and profile of the performed point operation.

| Algorithm 2: Montgomery ladder for point multiplication with random execution order [7] |
|---|
| **Input**: $P$, $m=\{m_{t-1}, ..., m_0\}$, $m_{t-1}=1$, random bits $r_{t-2}, ..., r_0$ |
| **Output**: $R=mP$ |
| 1. $R_0=P$ |
| 2. $R_1=2P$ |
| 3. **for** $i=t-2$ *to* $0$ **do** |
| 4.   **if** $m_i =1$ **then** |
| 5.     **if** $r_i =0$ **then** |
| 6.       $T_0=R_0+R_1$; $T_1=2R_1$; |
| 7.       $R_0=T_0$; $R_1=T_1$; |
| 8.     **else** |
| 9.       $T_1=2R_1$; $T_0= R_0+R_1$; |
| 10.       $R_0=T_0$; $R_1=T_1$; |
| 11.     **end** |
| 12.   **else** |
| 13.     **if** $r_i =0$ **then** |
| 14.       $T_1=R_0+R_1$; $T_0=2R_0$; |
| 15.       $R_0=T_0$; $R_1=T_1$; |
| 16.     **else** |
| 17.       $T_0=2R_0$; $T_1= R_0+R_1$; |
| 18.       $R_0=T_0$; $R_1=T_1$; |
| 19.     **end** |
| 20.   **end** |
| 21.   {$R_1$- $R_0$ *remains invariant*} |
| 22. **end** |
| 23. $R=R_0$ |

The design that executes the point operations in a randomized order in each iteration of the Montgomery ladder is considered to be more resistant against SCA attacks, especially against vertical DPA attacks which mainly target storing values in registers. We denoted this design further as *design with countermeasures* and the other one as *design without countermeasures* despite the fact that both designs used the complete addition law for calculating EC point additions and point doublings. The structure of the investigated designs is shown in Figure 1.

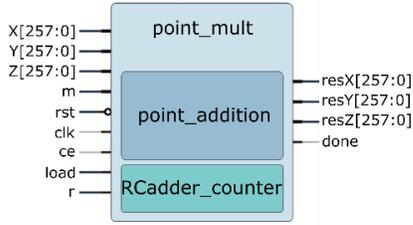

Figure 1. Structure of the investigated designs.

The designs are accelerators for the point multiplication $mP$ for the Elliptic Curve *secp256k1* [13]. Inputs of each design are: the scalar $m$; $X$, $Y$ and $Z$ coordinates of the EC point $P$, a set of control signals (*rst, clk, ce, load*) and a random number $r$. The *design without countermeasures* has the input $r$ but this value is not involved in any calculation. For the *design with countermeasures* a random bit value has to be provided to the input $r$ for a random execution order. Setting the *load* input to 1 on the rising clock edge will load input values into internal registers, and setting the *ce* input to 1 starts the point multiplication. The multiplication result is available at the outputs *resX*, *resY*, *resZ* after the *done* pin is set to active high.

The main block of the design is the *point_addition* module implementing the complete addition law. Its structure is given in Figure 2.

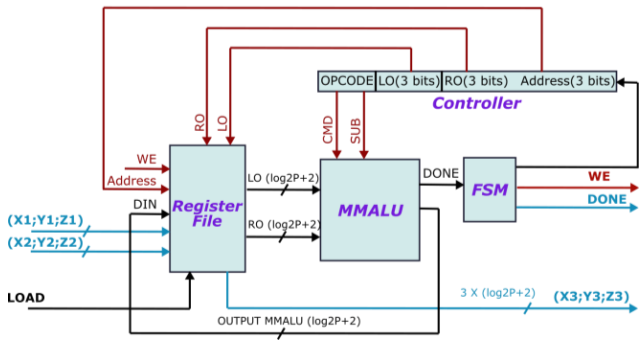

Figure 2. Structure of the *point_addition* module, corresponding to [8].

The *point_addition* module consists of the following main blocks:
- MMALU – Montgomery Modular Arithmetic Logic Unit. It allows to perform Montgomery modular addition, subtraction, or multiplication;
- Register File – 11 registers to store intermediate results. 6 registers are used as inputs, 3 of them are reused for the output and the remaining 5 are temporary registers;
- FSM – finite state machine, that ensures the sequence of operations to be executed;
- Controller – transfers an FSM opcode to control signals for the MMALU and the Register file.

More details on the designs' implementation are given in [9].

For the simulations discussed in this paper we used Cadence SimVision 15.20-s053. A part of the simulations of the *design with countermeasures* is shown in Figure 3. It can be seen that parameter $r_i$ (see signals *r_t* in Figure 3. ) determines the sequence of point operations, see signals *state* in Figure 3. after initialization (marked with time flag TimeA). For $r_i$ = '0' a point addition (*state=s_add*) is followed by point doubling (*state=s_double*). For $r_i$ = '1' a point doubling (see time flag TimeB, *state=s_double*) is followed by a point addition (*state=s_add*). The time for the processing of a single bit of the scalar $m$ can be also seen in our simulation in Figure 3. It is the time between time flags TimeA and TimeB. Processing of a single bit of the scalar $m$ takes TimeB-TimeA=*677675 ns – 226325 ns = 451350 ns*. It corresponds to 9027 clock cycles for a clock cycle period of 50 ns.

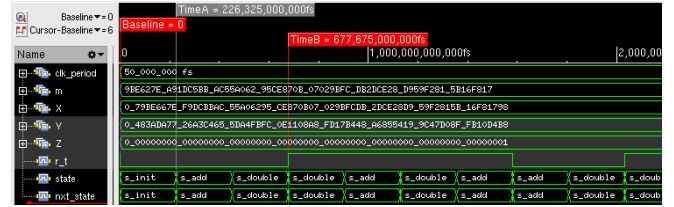

Figure 3. Beginning of the simulation for the *design with countermeasures* demonstrating a sequence of point addition and point doubling operations depending on parameter $r_i$ (see Algorithm 2).

### III. ANALYSIS OF THE SIMULTED TRACES

We synthesized both designs, i.e. the one without and the one with countermeasures, for the IHP 250 nm cell library SGB25V [14] using Synopsys Design Compiler Version K-2015.06-SP2. The maximum achieved clock frequency is 23.256 MHz (43 ns clock cycle period).

In our experiments we did not use the 20-bit long scalar $m$ = 0xbe9ff that was used in the original testbench as it is pretty short and predominantly consists of ones. Instead we used a 252-bit long random scalar:

$m$=0x9be627ea91dc5bbac55a06295ce870b07029bfcdb2dce28d959f2815b16f817

For the *design with countermeasures* we used the following scalar $r$ to ensure a random execution order:

$r$=3746cb5ed29e53453b0ff49f78e88bea61d8de75b8f55ab9a112d06bad0afc9.

We used the basis point of EC *secp256k1* [13] as the input point $P$ corresponding to the original testbench.

For our experiments we used a design synthesized for a frequency of 20 MHz. The area of the *design without countermeasures* is 1.933 mm$^2$. The area of the *design with countermeasures* is 2.364 mm$^2$, i.e. 22% larger than the one of the *design without countermeasures*. The difference in area for the *design with* and the *design without countermeasures* is caused by the additional temporary registers in the Montgomery ladder in the *design with countermeasures*. Parameters of the synthesized designs are shown in TABLE I. The designs synthesized for the NanGate 45 and IHP SGB25V cell libraries use almost the same number of gates expressed in gate equivalents (kGe).

TABLE I. DESIGN PARAMETERS

| Design | Synthesis library | Area | Clock Frequency, MHz | kP calculation time, ms |
|---|---|---|---|---|
| Without countermeasures | NanGate 45 | 66.51 kGe | 100 | 23.06 |
| | IHP 250 | 68.49 kGe | 20 | ~115 |
| With countermeasures | NanGate 45 | 81.89 kGe | 100 | 23.06 |
| | IHP 250 | 83.76 kGe | 20 | ~115 |

In order to test the resistance of both designs against horizontal attacks we generated power traces of the *kP* execution using Synopsys PrimeTime (R) Version Q-2019.12-SP1. Both designs require about 114 ms for processing a 252-bit long scalar.

## IV. COMPARISON TO THE MEAN ATTACK

In [5], [10] *the comparison to the mean* attack was introduced. It is based on the fact that the measured *kP* trace is a sequence of the processing of '1' and '0' values of the scalar *k*. We call the samples in the trace representing the processing of a single bit of the scalar a slot. Each trace can be split into two sets of slots, one representing the processing of '1' and the second one representing the processing of '0'. In an ideal case these two sets cannot be distinguished, i.e. their mean shapes $mean_0$ and $mean_1$ are pointwise (or sample-wise) equal. But normally there are differences that allow distinguishing the two sets. These differences may be found in all samples of the mean shapes or only in a few samples. If these differences are significant such a design is vulnerable to simple SCA. Otherwise an attacker can still try to distinguish the two sets, calculating a mean shape *mean* for the whole trace. The mean trace is then between both mean shapes: $mean_0 < mean < mean_1$ or $mean_0 > mean > mean_1$.

The attacker can determine key candidates by comparing the sample value with number *j* in the $i^{th}$ slot to the sample value with number *j* in the mean slot. If it is smaller then $k^{candidate\_j}_i = 1$, else the $i^{th}$ bit of the $j^{th}$ key candidate is equal to '0'. By applying this to all slots the attacker extracts *j* key candidates. Calculating $k^{candidate\_j} \cdot G$ (here *G* is the base point of the EC) and comparing the results with the public key of the attacked person the attacker can conclude if one of the extracted key candidates is equal to the processed scalar *k* or not. Evaluating the success of an attack is by far simpler for the designers, as they know the processed scalar *k* and can compare each of the key candidates bitwise with the scalar *k* and use this to calculate the success rate and to determine exactly which bits of the scalar were determined correctly.

We started with attacking the *design without countermeasures*. We run the attack described above analysing a trace simulated for the top module (*point_mult*) during the execution of the *mP* operation with *m, P* given in section III.

We evaluate the success of the attack by comparing the extracted key candidates with the real scalar *m* processed in the trace. We compare a key candidate with the scalar *m* bitwise and express the number of correctly extracted bits in per cent, denoted further as the correctness of the key candidates. Due to the fact that simulated traces are noise free we set the time step in our simulation equal to the period of the clock cycle. Thus, each clock cycle in our simulation was represented only with 1 sample. We decided to do so to avoid a huge memory consumption as the processing of a single key bit in the algorithm's main loop takes 9027 clock cycles, i.e. the complete trace is about 2.3 Mio. clock cycles long. The file in which the simulation results for a single top block are stored is about 0.5 GB even when a coarse time step for the simulation is used. In our attack we extracted 1 key candidate per clock cycle, i.e. we obtained 9027 key candidates. The success of our attack represented as the correctness of each of key candidates is shown in Figure 4.

In total there are three key candidates out of 9027 with a correctness of more than 75%. The key candidate, obtained using the first clock cycle of each slot, has a correctness of 100%. The two key candidates obtained for the clock cycles 4514 and 9027 have a correctness of 89.8 % and 92.9 % respectively.

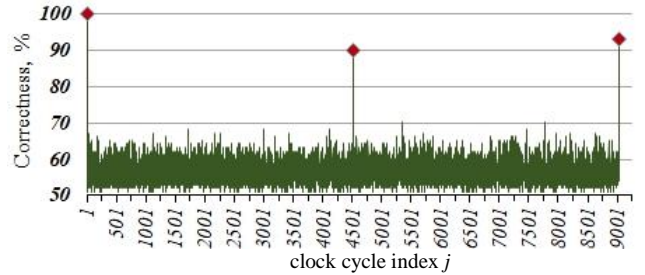

Figure 4. Results of our horizontal attack using *comparison to the mean* against the *design without countermeasures*.

We repeated the same attack using the simulated power trace for the top module of the *design with countermeasures*. The attack results are shown in Figure 5.

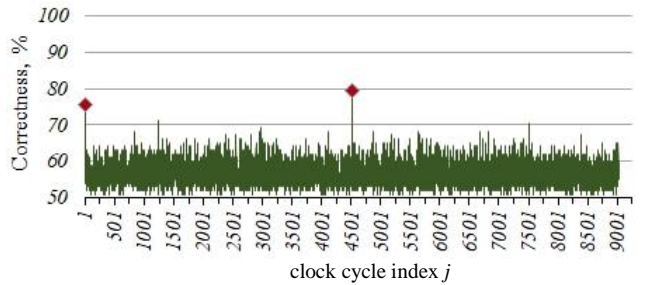

Figure 5. Results of our horizontal attack using *comparison to the mean* against the *design with countermeasures*.

The correctness of the key candidate derived from the first clock cycle dropped from 100 % down to 75.5%. The correctness of the key candidates derived from clock cycle 4514 and 9027 dropped from 89.8% to 79.6% and to less than 70%, respectively.

## V. IMPROVING OF ATTACK SUCCESS

The success rate of the performed horizontal DPA attack is high for both designs. Due to the fact that we attacked the trace simulated with a coarse simulation time step we decided to perform the simulation a second time, with 100 samples per clock cycle. The file in which the simulation results are stored is about 48 GB. The automated SPA attack was described in [15] allows easy to detect the observable differences in shapes of slots corresponding to the processing of different key bit values.

When running simple SCA, attackers apply the following principles:
- The shapes of the processing of key bit values '0' look similar to each other;
- The shapes of the processing of key bit values '1' look similar to each other;
- The shapes of the processing of key bit values '0' are distinguishable from the shapes of the processing of key bit values '1' using simple visual inspection;
- The analysed trace is a set of two different kinds of shapes, i.e. a sequence of '0'-shapes and '1'-shapes;
- The correctly extracted sequence of '0'- and '1'-shapes corresponds to the used secret i.e. to the scalar *m*.

Thus, for executing a simple analysis attack the attacker looks at the measured trace and tries to apply the above listed principles, i.e. the instruments used for the analysis are the eyes of the attacker and his natural intelligence. Due to this fact, the success of simple analysis attacks depends directly on the distinguishability of the '0'- and '1'-shapes. If the difference is significant and can be easy seen the sequence of the shapes in the trace is clear. The key candidate extracted using this evident sequence matches with the real processed key to 100% i.e. all bits of the key are revealed correctly.

We programmed the SCA procedure described above to distinguish shapes of '0' and '1' slots by comparing them sample-wise. Thus, our program helps to detect very small differences in a similar way as a magnifying glass would do.

We applied the automated SPA attack to both designs i.e. to the design without and the one with countermeasures.

### A. Design without countermeasures

We attacked the power trace of the design's top module simulated with a sampling rate of 100 samples per clock cycle. Please note that in this case each slot consists of 9027·100=902700 samples. For the design *without countermeasures* we obtained 8 samples that allowed revealing the processed scalar successfully:
- 2 samples in clock cycle 1,
- 2 samples in clock cycle 2,
- 2 in clock cycle 4514,
- 2 in clock cycle 9027,

i.e. we obtained more key candidates that were 100% correct than when analysing the coarse simulated trace. Please note that the correctness of the best key candidates for clock cycle 1 is about 90%. The analysis of the coarse simulated trace indicated clock cycle 1 as the source of the SCA leakage. The coarse sampling rate during the simulations is a kind compressing the trace. The value representing a clock cycle is the mean power consumption for this clock cycle and can be calculated as arithmetical mean of all samples in the clock cycle if a fine simulation time step is used. The difference in the correctness of the key candidates clearly shows that compression influences the success of the attack even for noise free traces.

Our tool (the automated SPA attack) provides only those sample numbers for which the key was revealed with a correctness of 100%. It does not provide any information on the correctness of key candidates extracted for other samples. So in order to learn the correctness of the key candidates derived from other samples we run a horizontal DPA sample-wise to determine the correctness of all key candidates. We attacked the traces simulated with the fine simulation time using our horizontal DPA sample-wise. We obtained 902700 key candidates for each attacked design. The results of the attack against the *design without countermeasures* are shown in Figure 6. In Figure 6. -*(a)* the correctness of all key candidates is shown as a line. Figure 6. -*(b)*, Figure 6. -*(c)* and Figure 6. -*(d)* show parts of the Figure 6. -*(a)* corresponding to the key candidates with a high correctness zoomed in.

In total there are 5 clock cycles each of them containing 2 points for which the correctness of the extracted key candidates is higher than 90%. 8 of these 10 points correspond to key candidates with a correctness of 100%.

In order to determine which of the design blocks is the source of the leakage, we repeated the attack against the main blocks and sub-blocks of the designs i.e. we analysed the power traces simulated for *RCadder_counter*, *point_addition* module (see Figure 1. ) as well as for the internal blocks of the *point_addition* module shown in Figure 2. i.e. Register file, MMALU and FSM.

The main source of the leakage is the Register file block. Analysing the power trace of this block reveals the processed scalar with a correctness of 100% in points of the clock cycles 1, 2, 4514 and 4515. The analysis of its parent block – the *point_addition* module – shows that the correctness of the keys revealed remains at the level of 100% for the clock cycles 1, 2 and 4514 whereas it is slightly reduced down to 97.9 % in clock cycle 4515.

Another leakage source was discovered in clock cycle 9027 of the top block *point_mult*. Here the processed scalar can be revealed with a correctness of 100% as well. The possibility to reveal the key in clock cycles 1, 2, 4514 and 4515 originates from the block Register file as discussed earlier.

The remaining blocks of the design do not exhibit any significant leakage, as the scalar can be revealed with maximum correctness of 56.1% for the *RCadder_counter*, 71.4% for the MMALU block, 61.2% and 53.1% for the FSM and Controller blocks, respectively.

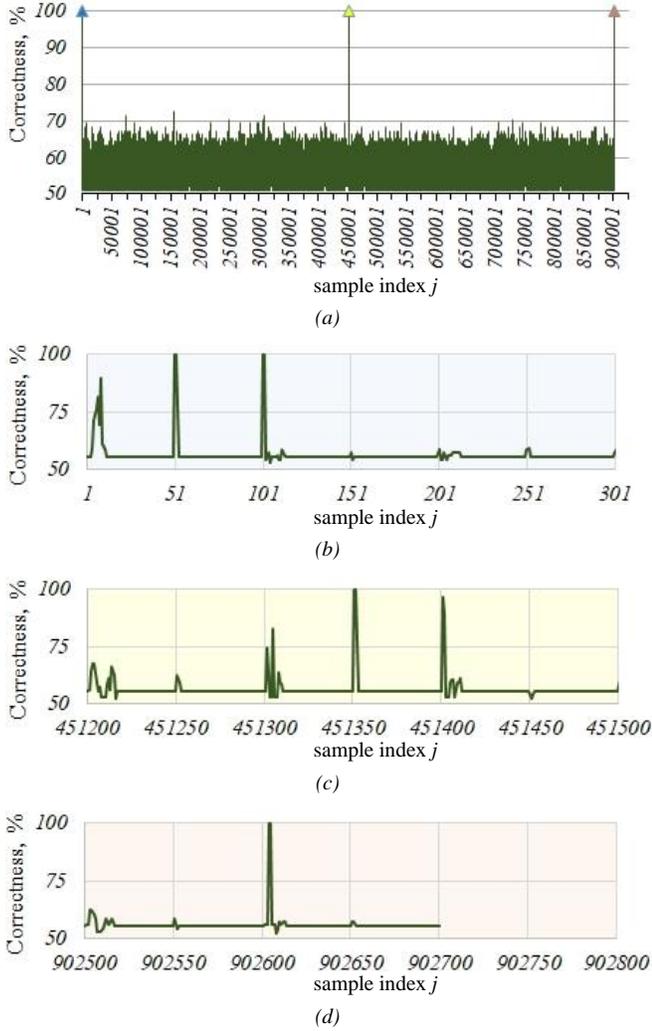

Figure 6. Results of the sample-wise horizontal DPA attack against the *design without countermeasures (a)* and fragments corresponding to the key candidates with a high correctness, zoomed in *(b)-(d)*.

TABLE II. presents a list of the clock cycles corresponding to the strong leakage sources that cause the key to be revealed with a correctness of 100%.

TABLE II. ATTACK RESULTS FOR MAIN BLOCKS OF THE DESIGN WITHOUT COUNTERMEASURES

| Block name | Clock cycles corresponding to the strong leakage sources: the correctness of the extracted key candidates is 100% |
|---|---|
| *point_mult* (top block) | 1, 2, 4514, 4515, 9027 |
| ↳ *point_addtion* | 1, 2, 4514, 4515 |
| ↳ Register file | 1, 2, 4514, 4515 |

### B. Design with countermeasures

We performed all attacks described earlier in this section against *design with countermeasures* analysing its power trace simulated with the fine sampling rate of 100 samples per clock cycle. Results of attacking the top block are presented in Figure 7. The correctness of the key revealed at the beginning of the slot is only 78.6% (see Figure 7. *-b)*) whereas the one of the *design without countermeasures* is 100%. The leakage at the end of the slot that was caused by the block *point_mult* disappeared completely. Only the leakage detected at points in the middle of the slot – samples 451351 and 451352, see Figure 7. *-c)* – still allows to fully reveal the processed scalar.

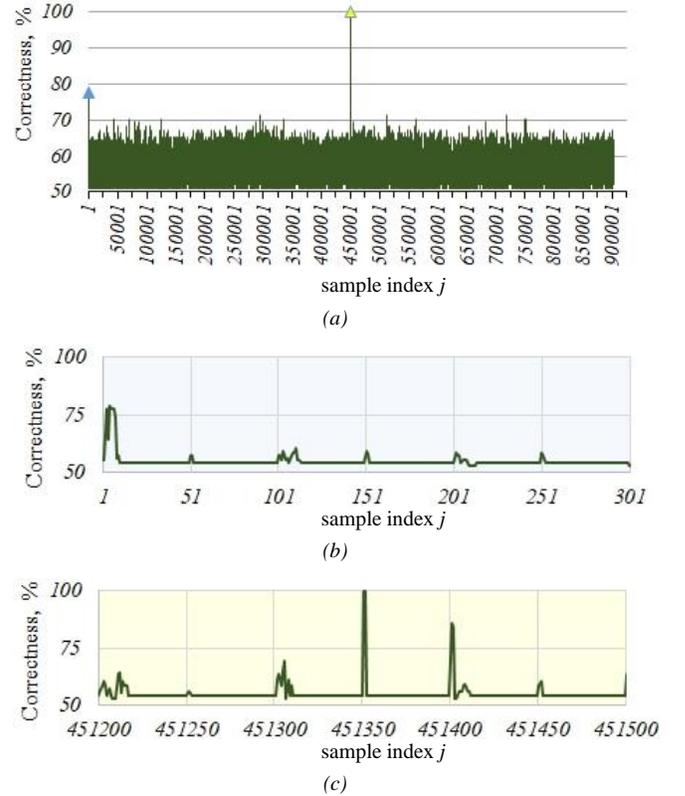

Figure 7. Results of the sample-wise horizontal DPA attack against the *design with countermeasures (a)* and fragments corresponding to the key candidates with a high correctness, zoomed in *(b), (c)*.

We also attacked the individual blocks of the design. The results are presented in TABLE III.

TABLE III. ATTACK RESULTS FOR THE DESIGN WITH COUNTERMEASURES AND ITS MAIN BLOCKS

| Block name | Clock cycle (correctness) |
|---|---|
| *point_mult* (top block) | 1 (78.6%), 4514 (100%), 4515 (85.7%) |
| ↳ *point_addtion* | 1 (79.6%), 4514 (100%), 4515 (84.7%) |
| ↳ Register file | 1 (79.6%), 4514 (100%), 4515 (100%) |

As in the case of the *design without countermeasures*, the source of the leakage is the block Register file that is an internal block of the *point_addition* module. The maximum correctness of the scalar revealed is about 60% for the blocks *RCadder_counter*, FSM and Controller. The best key candidate for the MMALU block has a correctness of 72.4%.

We use the sample 451352 of each slot, that corresponds to the strong leakage source, to visualize our automated SPA attack. Figure 8. shows the power consumed at that point in time during the processing of the first 50 bits of the scalar "9BE627EA91DC5" except of its most significant bit. It can be clearly seen if a bit equal to '0' is processed the power consumed is higher than 1.2 mW.

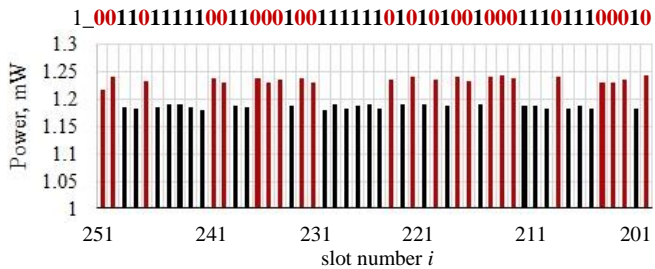

Figure 8. Values of the power consumed in slot's sample 451352 extracted from each of the first 50 slots of the simulated power trace of the *point_mult* block.

## VI. CONCLUSIONS

In this paper we evaluated the resistance of two open source ECC designs against selected horizontal attacks. One design is an implementation of the Montgomery ladder. The second one is an implementation of the Montgomery ladder with a randomized sequence of EC point operations. Both designs perform EC point addition and EC point doubling corresponding to the complete addition formulas that ensure that power profiles of the EC point operations look very similar. In the literature the Montgomery *kP* algorithm is reported as resistant against simple SCA due to the fact that the sequence of operations in its main loop is the same for the processing of each key bit. Randomizing the sequence of the EC point operations in the main loop of the algorithm can countermeasure vertical data bit differential SCA attacks. The same holds true for vertical and horizontal address bit differential attacks, especially if the profiles of the EC point operations are indistinguishable. We performed two horizontal attacks: a DPA using *the comparison to the mean* for the analysis of simulated power traces (with coarse and fine simulation time steps) and an automated SPA. All attacks were successful, i.e. some of the extracted key candidates were equal to the processed scalar. By attacking the traces of internal blocks of the designs we determined that the activity of the Register file block is the main SCA leakage source. The analyzed simulated traces are noise free and only a few points in the trace caused the strong SCA leakage. Therefore, the countermeasure based on the randomized sequence of point operations and the use of the universalized point addition can probably offer some kind of protection against a broad spectrum of SCA attacks. It is, however, advisable to combine this countermeasure with additional noise sources for selected clock cycles or with additional protection mechanisms against horizontal address bit DPA [16].